\documentstyle[sprocl]{article}

\input{psfig}

\newcommand{\nc}{\newcommand}
\nc{\renc}{\renewcommand}
\nc{\bort}[1]{}
\nc{\pub}[4]{{\it #1}{\bf #2} {#3}{(#4)}.}
\nc{\aap}[3]{{\it  Astron.\ Astrophys.\ }{{\bf #1} {(#3)} {#2}}}
\nc{\advp}[3]{{\it  Adv.\ in\ Phys.\ }{{\bf #1} {(#3)} {#2}}}
\nc{\annp}[3]{{\it  Ann.\ Phys.\ (N.Y.)\ }{{\bf #1} {(#3)} {#2}}}
	\nc{\annraa}[3]{{\it Ann.\ Rev.\ Astron.\ Astrophys.\ }{{\bf #1} {(#3)} {#2}}}
\nc{\apl}[3]{{\it  Appl. Phys. Lett. }{{\bf #1} {(#3)} {#2}}}
\nc{\apj}[3]{{\it  Ap.\ J.\ }{{\bf #1} {(#3)} {#2}}}
\nc{\apjl}[3]{{\it  Ap.\ J.\ Lett.\ }{{\bf #1} {(#3)} {#2}}}
\nc{\app}[3]{{\it Astropart.\ Phys.\ }{{\bf #1} {(#3)} {#2}}}

\nc{\cmp}[3]{{\it  Comm.\ Math.\ Phys.\ }{{ \bf #1} {(#2)} {#3}}}
\nc{\cqg}[3]{{\it  Class.\ Quant.\ Grav.\ }{{\bf #1} {(#3)} {#2}}}
\nc{\epl}[3]{{\it  Europhys.\ Lett.\ }{{\bf #1} {(#3)} {#2}}}
\nc{\ijmp}[3]{{\it Int.\ J.\ Mod.\ Phys.\ }{{\bf #1} {(#3)} {#2}}}
\nc{\ijtp}[3]{{\it Int.\ J.\ Theor.\ Phys.\ }{{\bf #1} {(#3)} {#2}}}
\nc{\jmp}[3]{{\it  J.\ Math.\ Phys.\ }{{ \bf #1} {(#2)} {#3}}}
\nc{\jpa}[3]{{\it  J.\ Phys.\ A\ }{{\bf #1} {(#3)} {#2}}}
\nc{\jpc}[3]{{\it  J.\ Phys.\ C\ }{{\bf #1} {(#3)} {#2}}}
\nc{\jpg}[3]{{\it J.~Phys.~G:~Nucl.~Part.~Phys.~}{{\bf #1} {(#2)}{#3}}}
\nc{\jap}[3]{{\it J.\ Appl.\ Phys.\ }{{\bf #1} {(#3)} {#2}}}
\nc{\jpsj}[3]{{\it J.\ Phys.\ Soc.\ Japan\ }{{\bf #1} {(#3)} {#2}}}
\nc{\kdmfm}[3]{{\it Kong.\ Dan.\ Mat.\ Fys.\ Med.\ }{{\bf #1} {(#3)} {#2}}}
\nc{\lmp}[3]{{\it Lett.\ Math.\ Phys.\ }{{\bf #1} {(#3)} {#2}}}
\nc{\lncim}[3]{{\it  Lett.\ Nuov.\ Cim.\ }{{\bf #1} {(#3)} {#2}}}
\nc{\mpl}[3]{{\it  Mod.\ Phys.\ Lett.\ }{{\bf #1} {(#3)} {#2}}}
\nc{\naturw}[3]{{\it  Naturwiss.\ }{{\bf #1} {(#3)} {#2}}}
\nc{\ncim}[3]{{\it  Nuov.\ Cim.\ }{{\bf #1} {(#3)} {#2}}}
\nc{\np}[3]{{\it  Nucl.\ Phys.\ }{{\bf #1} {(#3)} {#2}}}
\def\npfs#1#2#3#4{{\it  Nucl.\ Phys.\ }{{\bf #1 [FS#2]} {(#3)} {#4}}}
\nc{\pr}[3]{{\it Phys.\ Rev.\ }{{\bf #1} {(#3)} {#2}}}
\nc{\pra}[3]{{\it  Phys.\ Rev.\ }{{\bf A#1} {(#2)} {#3}}}
\nc{\prb}[3]{{\it  Phys.\ Rev.\ }{{\bf B#1} {(#2)} {#3}}}
\nc{\prc}[3]{{\it  Phys.\ Rev.\ }{{\bf C#1} {(#2)} {#3}}}
\nc{\prd}[3]{{\it  Phys.\ Rev.\ }{{\bf D#1} {(#2)} {#3}}}
\nc{\prl}[3]{{\it Phys.\ Rev.\ Lett.\ }{{\bf #1} {(#3)} {#2}}}
\nc{\pl}[3]{{\it  Phys.\ Lett.\ }{{\bf #1} {(#3)} {#2}}}
\nc{\prep}[3]{{\it Phys.\ Rep.\ }{{\bf #1} {(#3)} {#2}}}
\nc{\prsl}[3]{{\it Proc.\ R.\ Soc.\ London\ }{{\bf #1} {(#3)} {#2}}}
\nc{\ptp}[3]{{\it  Prog.\ Theor.\ Phys.\ }{{\bf #1} {(#3)} {#2}}}
\nc{\ptps}[3]{{\it  Prog.\ Theor.\ Phys.\ suppl.\ }{{\bf #1} {(#3)} {#2}}}
\nc{\physa}[3]{{\it  Physica\ A\ }{{\bf #1} {(#3)} {#2}}}
\nc{\physb}[3]{{\it  Physica\ B\ }{{\bf #1} {(#3)} {#2}}}
\nc{\phys}[3]{{\it Physica\ }{{\bf #1} {(#3)} {#2}}}
\nc{\rmp}[3]{{\it  Rev.\ Mod.\ Phys.\ }{{\bf #1} {(#3)} {#2}}}
\nc{\rpp}[3]{{\it Rep.\ Prog.\ Phys.\ }{{\bf #1} {(#3)} {#2}}}
\nc{\sjnp}[3]{{\it Sov.\ J.\ Nucl.\ Phys.\ }{{\bf #1} {(#3)} {#2}}}
\nc{\spjetp}[3]{{\it Sov.\ Phys.\ JETP\ }{{\bf #1} {(#3)} {#2}}}
\nc{\yf}[3]{{\it Yad.\ Fiz.\ }{{\bf #1} {(#3)} {#2}}}
\nc{\zetp}[3]{{\it Zh.\ Eksp.\ Teor.\ Fiz.\ }{{\bf #1} {(#3)} {#2}}}
\nc{\zp}[3]{{\it Z.\ Phys.\ }{{\bf #1} {(#3)} {#2}}}
\nc{\zpc}[3]{{\it Z.\ Phys.\ C\ }{{\bf #1} {(#3)} {#2}}}
\nc{\ibid}[3]{{\sl ibid.\ }{{\bf #1} {#2} {#3}}}
\newlength{\overeqskip}
\newlength{\undereqskip}
\def\be{\begin{equation}}
\def\bea{\begin{eqnarray}}
\nc{\eea}{\vspace{\undereqskip}\end{eqnarray}}
\nc{\ee}{\vspace{\undereqskip}\end{equation}}
\nc{\bdm}{\begin{displaymath}}
\nc{\edm}{\end{displaymath}}
\nc{\dpsty}{\displaystyle}
\nc{\bc}{\begin{center}}
\nc{\ec}{\end{center}}
\nc{\ba}{\begin{array}}
\nc{\ea}{\end{array}}
\nc{\bab}{\begin{abstract}}
\nc{\eab}{\end{abstract}}
\nc{\btab}{\begin{tabular}}
\nc{\etab}{\end{tabular}}
\nc{\bit}{\begin{itemize}}
\nc{\eit}{\end{itemize}}
\nc{\ben}{\begin{enumerate}}
\nc{\een}{\end{enumerate}}

\nc{\arreq}{&\!\!\!=\!\!\!&}
\nc{\arrmi}{&\!\!\!!-\!\!\!&}
\nc{\arrpl}{&\!\!\!+\!\!\!&}
\nc{\arrap}{&\!\!\!\approx\!\!\!&}
\nc{\non}{\nonumber}
\nc{\nn}{\nonumber\\}
\nc{\align}{\!\!\!\!\!\!\!\!&&}

\nc{\mat}[4]{{\left(\ba{cc} #1 & #2 \\ #3 & #4 \ea\right)}}

\def\simleq{\; \raise0.3ex\hbox{$<$\kern-0.75em
      \raise-1.1ex\hbox{$\sim$}}\; }
\def\simgeq{\; \raise0.3ex\hbox{$>$\kern-0.75em
      \raise-1.1ex\hbox{$\sim$}}\; }

\nc{\DOT}{\hspace{-0.08in}{\bf .}\hspace{0.1in}}
\nc{\Laada}{\hbox {$\sqcap$ \kern -1em $\sqcup$}}
\nc\loota{{\scriptstyle\sqcap\kern-0.55em\hbox{$\scriptstyle\sqcup$}}}
\nc\Loota{{\sqcap\kern-0.65em\hbox{$\sqcup$}}}
\nc\laada{\Loota}
\nc{\qed}{\hskip 3em \hbox{\BOX} \vskip 2ex}
\def\Re{{\rm Re}\hskip2pt}

\nc{\real}{{\rm I \! R}}
\nc{\Z}{{\sf Z \!\!\! Z}}
\nc{\complex}{{\rm C\!\!\! {\sf I}\,\,}}

\def\bigid{\leavevmode\hbox{\small1\kern-3.8pt\normalsize1}}
\def\id{\leavevmode\hbox{\small1\kern-3.3pt\normalsize1}}
\nc{\slask}{\hspace{0.1em}\not\hspace{-0.25em}}
\nc{\bis}{{\prime\prime}}
\nc{\pa}{\partial}
\nc{\na}{\nabla}
\def\>{\rangle}
\def\<{\langle}

\nc{\goto}{\rightarrow}
\nc{\swap}{\leftrightarrow}
\nc{\EE}[1]{ \mbox{$\times 10^{#1}$} }
\nc{\abs}[1]{\left|#1\right|}
\nc{\at}[2]{\left.#1\right|_{#2}}
\nc{\norm}[1]{\|#1\|}
\nc{\abscut}[2]{\abs{#1}_{\scriptscriptstyle#2}}
\nc{\vek}[1]{\hbox{\boldmath$#1$}}
\nc{\integral}[2]{\int\limits_{#1}^{#2}}

\nc{\dd}[2]{{{\partial #1}\over{\partial #2}}}
\nc{\ddd}[2]{{{{\partial}^2 #1}\over{\partial {#2}^2}}}
\nc{\dddd}[3]{{{{\partial}^2 #1}\over
        {\partial #2 \partial #3}}}
\nc{\dder}[2]{{{d #1}\over{d #2}}}
\nc{\ddder}[2]{{{d^2 #1}\over{d {#2}^2}}}
\nc{\dddder}[3]{{d^2 #1}\over
        {d #2 d #3}}

\nc{\cc}{\mbox{$c.c.$ }}
\nc{\hc}{\mbox{$h.c.$ }}
\nc{\cf}{cf.\ }
\nc{\erfc}{{\rm erfc}}
\nc{\Tr}{{\rm Tr\,}}
\nc{\tr}{{\rm tr\,}}
\nc{\pol}{{\rm pol}}
\nc{\sign}{{\rm sign}}
\nc{\bfT}{{\bf T }}

\nc{\cA}{{\cal A}}
\nc{\cB}{{\cal B}}
\nc{\cD}{{\cal D}}
\nc{\cE}{{\cal E}}
\nc{\cF}{{\cal F}}
\nc{\cG}{{\cal G}}
\nc{\cH}{{\cal H}}
\nc{\cL}{{\cal L}}
\nc{\cM}{{\cal M}}
\nc{\cO}{{\cal O}}
\nc{\cP}{{\cal P}}
\nc{\cQ}{{\cal Q}}
\nc{\cT}{{\cal T}}

\nc{\1}{\aa}
\nc{\2}{\"{a}}
\nc{\3}{\"{o}}
\nc{\4}{\AA}
\nc{\5}{\"{A}}
\nc{\6}{\"{O}}
\nc{\al}{\alpha}
\nc{\Del}{\Delta}
\nc{\e}{\epsilon}
\nc{\eps}{\epsilon}
\nc{\g}{\gamma}
\nc{\lam}{\lambda}
\nc{\om}{\omega}
\nc{\Om}{\Omega}
\nc{\ve}{\varepsilon}
\nc{\mn}{{\mu\nu}}
\nc{\ka}{\kappa}

\nc{\vph}{\varphi}

\begin{document}

\title{STABILIZATION OF CHROMOMAGNETIC FIELDS \\
 AT HIGH TEMPERATURE?}

\author{DAVID PERSSON}

\address{Department of  Physics and Astronomy\\ The University of British 
Columbia\\Vancouver, B.C. V6T 1Z1, Canada\\E-mail:  
persson@theory.physics.ubc.ca}


\maketitle\abstracts{ It is well known that a tachyonic mode appears in the
spectrum of Yang---Mills theory with a static uniform magnetic field,
 and that the free energy has an (unstable) minimum at finite magnetic field.
It is argued that spontaneous generation of magnetic field does not take place
 at high temperature due to nonperturbative magnetic screening.
Furthermore, the dispersion relation for gauge field fluctuations in an 
external magnetic field at high temperature is solved. The lowest energy mode
 is stable against spontaneous generation of magnetic fields since it acquires
 a thermal mass.
However, the resummed free energy (by necessity computed in the imaginary
time formalism) still shows an instability, unaffected by the resummation,
since the self-energy is vanishing at static Matsubara frequency.}

\section{The Instability and how it is Screened at High Temperature}
For non-abelian gauge theories there are classes of gauge equivalent potentials
corresponding to the same field strength. We shall in this letter only consider
the abelian like class, that is the only one reasonable in an early universe
scenario, and for simplicity only $SU(N=2)$.
 Let us therefore consider a static uniform (chromo-) magnetic field
in $z$ direction in space, with the potential
$A^a_\mu=\delta^{a3}(0,0,-Bx,0)$,
where the field points in the 3-direction in group space.
The $SU(2)$ Lagrangian with this background field is then rewritten 
in terms of the charged Vector Field  $W_\mu=(W^1_\mu+iW^2_\mu)/\sqrt{2}$.
The energy spectrum reads
\be
	E^2(k_z,l,\sigma)=k_z^2+(2l+1)gB-2\sigma gB~~,
\ee
where the term  $(2l+1)gB$, $l=0,1,2,\ldots$,
comes from the orbital angular momentum and the term $2\sigma
gB$, $\sigma=\pm 1$, is the spin energy. The momentum parallell to $B$ is 
$k_z$.
Obviously, we have an instability in the lowest Landau level (LLL), $l=0,~
\sigma=1$.
At $T=0$, this leads to a spontaneous generation of a magnetic 
field~\cite{Savvidy77}
$gB_0=\Lambda^2=\lambda^2 \exp\left( -\frac{48 \pi^2}{11N g^2(\lambda)}
\right)$, where $\lambda$ is the renormalization scale, and $g^2(\lambda)$ is
running so that $\Lambda$ is independent of $\lambda$. However,
the free energy acquires an imaginary part,
 so this minimum is unstable~\cite{NielsenO78},
A possible groud-state has to be varying over the non-perturbative scale
$1/\Lambda$---``Copenhagen Vacuum''~\cite{AmbjornNO79}.

\smallskip

At high temperature, we have for an asymptotically free theory the following
hierarchy of scales (in due order)

\bit
\item[$T$:]
The  temperature is the typical energy of particles in the plasma and the
inter-particle distance is $\sim 1/T$.
\item[$gT$:]
The interaction of soft particles ($p\sim gT$) with hard particles ($p\sim
T$) generates a thermal mass of order $gT$. Static electric (but not magnetic)
fields are screened over the length scale $1/gT$.
\item[$g^2T$:]
On this momentum scale Yang-Mills theory becomes non-perturbative.
 Theoretical arguments and 
 lattice simulations show that non-abelian magnetic
fields are screened over the length scale $1/g^2T$. 
\item[$\Lambda$:]
The strong coupling scale below which the vacuum theory becomes
non-perturbative.
\eit
 
Due to the non-perturbative magnetic screening the extension of the LLL 
is much larger than the length-scale over which the magnetic field can be
 constant $1/\Lambda \gg 1/g^2T$. The would be unstable mode thus will not see
a uniform field, and the Saviddy mechanism for spontaneous generation of
magnetic field cannot operate.

\section{External Chromomagnetic Fields}	\label{sec-ext}
\begin{figure}[t]
\centerline{\psfig{figure=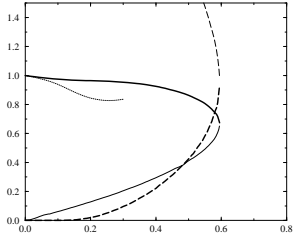,height=1.5in}}
\caption{The solid lines show the two branches of the solution to
the dispersion relation for $k_z=0$ (fat---stable, thin---unstable). 
The dashed lines show
the corresponding imaginary parts, and the dotted line is the spectral 
weight of the stable mode.  \label{fig:dr}}
\end{figure}

If we instead
assume an external magnetic field, generated by some other mechanism, we may
consider the  hierarchy of scales:
\be
	T^2 \gg\cM^2\equiv \frac{N}9 (gT)^2 \gg gB \gg (g^2T)^2~~~.
\ee
In order to investigate if the instability is screened at high temperature,
we now need to consider the dispersion relation obtained from the effective
Lagrangian
\bea
\lefteqn{
\int d^4x\,d^4x'\,W_\mu^\dagger(\kappa,x)[-\delta(x-x')
	\left(g^{\mu\nu}D^2-D^\mu D^\nu -2igF^{\mu\nu}\right)}~~~~~~~~~~~~~~~~~
	\nonumber \\
	&& \left. -\Pi^{\mu\nu}(x,x')\right]W_\nu(\kappa',x'
	)~~~,~~~~~~~~~~~~~~~~~~~~~~~~~~~~~~~~~
\eea
where $\Pi$ is the gluon self-energy in a magnetic field at high temperature.
With wave-functions corresponding to the LLL, unstable at $T=0$, the 
dispersion relation reads in momentum space
\be	\label{LLLPi}
	k_0^2 +gB -k_z^2 + \Pi_{\rm LLL}(k_0,k_z)=0~~~,
\ee
where we integrate over perpendicular momenta
\be \label{perpint}
	\Pi_{\rm LLL}(k_0,k_z)=\int_0^\infty \frac{2p_\perp\,dp_\perp}{gB} 
	e^{-p_\perp^2/gB} w_\mu^{\rm LLL} \Pi_{\rm HTL}^{\mu\nu}(k_0,k_z,
	p_\perp) w_{\nu}^{\rm LLL}~~~,
\ee
and $w_\mu$ is the polarization vector in LLL. It turns out that the leading
high temperature correction comes only from the ordinary hard thermal loop
approximation of the gluon self-energy tensor, conveniently separated in
its longitudinal ($\Pi_{\rm L}$), and transverse ($\Pi_{\rm T}$) parts.
To leading order,  the magnetic field only enters through the external states
being Landau levels.
In order to investigate the instability, let us consider $k_z=0$. For
$k_0 \sim \cM \gg \sqrt{gB}$, we find to leading order
\be
	k_0^2+gB-\cM^2\left(1+\frac{2gB}{5\cM^2}\right)=0~~,
\ee
i.e. $k_0\simeq \cM(1-3gB/10\cM^2)$, a stable solution. However, if we instead
expand for small $k_0$, we get to leading order
\be
	k_0^2+gB+i\frac{3\pi^{3/2}}{8}\frac{k_0\cM^2}{\sqrt{gB}}=0~~,
\ee
with the leading solution $k_0 \simeq i 16/(3 \pi^{3/2})\, (gB)^{3/2}/\cM^2$.
This may thus be the signal of an instability. In Figure~\ref{fig:dr}, we
show the two branches obtained by solving the real part of the dispersion
relation for real $k_0$, and then determine the imaginary part on this solution.
Furthermore, the spectral weight of the stable mode is shown. For weak
magnetic fields it is close to unity, leaving no phase space for the
unstable mode.
However, work in progress shows that when considering the full dispersion 
relation, there is a branch with purely imaginary $k_0$, indicating the
survival of the instability at high temperatures. 

\section{The Resummed Free Energy}

 In order to avoid the tree level instability, it is necessary
to consider the resummed free energy, including the leading self-energy term,
as well as the corresponding counter term. However, this corresponds
to a sum over diagrams with vanishing momenta on the external lines (here
rather no external lines), so that the real time formalism is not
 straightforwardly applicable~\cite{EvansP95}.
Let us neglect the spatial momenta, and 
first consider the self-energy like function in imaginary time
\be
	\Theta_{\rm IT}=-T\sum_n\Delta(i\om_n,k)=T\sum_n 
	\frac1{\om_n^2+k^2+\Pi(i\om_n,k)}~;~~~\om_n=2\pi n T~~~,
\ee
where the self-energy $\Pi$, has a branch cut along the imaginary axis.
This equals the real time expression
\be
	\Theta_{\rm IT}=\Theta_{\rm RT}=\int \frac{dk_0}{2\pi} \varrho(k_0)
	\left\{ \frac12 + f_{\rm B}(k_0) \right\}~;~~ f_{\rm B}(k_0)=
	\frac1{e^{\beta |k_0 |-1}}~~~,
\ee
if we only have poles in $\Delta$ along the real axis. The spectral density is
\be
	 -i \ve(k_0) \varrho(k_0)={\rm Disc}\Delta(k_0)=
	\Delta(k_0+i\e)-\Delta(k_0-i\e)~, \e \rightarrow 0^+~~~,
\ee
and $\ve$ is the sign function. From this we may derive that  the partition 
function like quantity
\be
	\Psi_{\rm IT}=-\frac{T}2 \sum_n \ln\left\{ \beta^2\left[ \om_n^2 +k^2
	+\Pi(i\om_n,k) \right] \right\}~~~,
\ee
rewritten as an integral over real energies, takes the form
\bea
	\Psi_{\rm IT}&=&-\frac{T}2 \int \frac{dk_0}{2\pi} \left\{ 
	\frac{\beta |k_0|}2 +\ln\left(1-e^{-\beta |k_0|} \right) \right\} 
	2 |k_0| \left[ (1-\Re \nu)\varrho \right. \nn
	&&\left. -i \Re \Delta {\rm Disc}\, \nu \right] 
	-T \ln\left(1-e^{-\beta k} \right)~~~.
\eea
Obviously this differs from the correspondning real-time expression, that thus
 has to be wrong. It is thus necessary to compute the resummed free energy
 in the  imaginary time formalism. The contribution from LLL reads
\be
	\frac1{\beta V} \ln Z_{\rm LLL}= -\frac{gB}{(2\pi)^2} \int dk_z\,
	\sum_n \ln \left\{ \beta^2 \left[ \om_n^2+k_z^2-gB -
	\Pi_{\rm LLL}(k_0,k_z) \right] \right\}~~~.
\ee
The most infra-red sensitive part is for static Matsubara frequency
$i\om_{n=0}=0$, in which case the self-energy reads
\be
	\Pi_{\rm LLL}(k_0=0,k_z)=0~~~.
\ee
When computing the free energy, the instability thus appears to be unaffected
by the resummation. The instability may be removed by introducing a 
non-vanishing Polyakov loop\cite{MeisingerO97}. This essentially amounts
to replacing $\om_n \rightarrow \om_n-\phi/\beta$, and $\phi$ is determined by
minimizing the free energy.  Confer the recent studies of an external
 magnetic field on the lattice~\cite{KajantieLPRS98}, that did not show
any signs of the unstable phase, and the final remarks in 
Section~\ref{sec-ext}. It is the present author's definite
 opinion that this subject requires  further investigation.

\section*{Acknowledgments}
The first two thirds of this work was done in cooperation with Per Elmfors, as
reported in Ref.\cite{ElmforsP99}, to which we refer for further details and 
references. I would like to thank the organisers of SEWM `98. Financial 
support was obtained from the 
Nordic project ``Hot Non-Perturbative Particle Physics''; and NorFA under 
grant no. 96.15.053-O. The author's research was funded by STINT under 
contract 97/121.

\end{document}